\documentclass[runningheads]{llncs}
\usepackage[T1]{fontenc}
\usepackage{graphicx}
\usepackage{amsmath}
\usepackage{amssymb}
\begin{document}
\title{Adaptive Base Representation Theorem: An Alternative to Binary Number System}

\titlerunning{Adaptive Base Representation Theorem}

\author{Ravin Kumar \orcidID{0000-0002-3416-2679}}
\authorrunning{Ravin Kumar}
\institute{Department of  Computer Science, Meerut Institute of Engineering and Technology, Meerut-250005, Uttar Pradesh, India \\
\email{ravin.kumar.cs.2013@miet.ac.in}}
\maketitle              % typeset the header of the contribution
\begin{abstract}
This paper introduces the Adaptive Base Representation (ABR) Theorem and proposes a novel number system that offers a structured alternative to the binary number system for digital computers. The ABR number system enables each decimal number to be represented uniquely and using the same number of bits, $n$, as the binary encoding. Theoretical foundations and mathematical formulations demonstrate that ABR can encode the same integer range as binary, validating its potential as a viable alternative. Additionally, the ABR number system is compatible with existing data compression algorithms like Huffman coding and arithmetic coding, as well as error detection and correction mechanisms such as Hamming codes. We further explore practical applications, including digital steganography, to illustrate the utility of ABR in information theory and digital encoding, suggesting that the ABR number system could inspire new approaches in digital data representation and computational design.
\keywords{Number system \and Encoding \and Information theory \and Steganography \and Set theory \and Theorem.}
\end{abstract}
\section{Introduction}
The binary number system forms the foundation of modern digital computing, widely favoured for its simplicity, reliability, and storage efficiency. Binary representation is integral to various aspects of digital systems, from data storage and processing to encoding and transmission. However, as digital applications evolve, exploring alternative number systems presents promising opportunities to address unique computational challenges and inspire new approaches in data representation. This paper proposes a novel number system, termed the Adaptive Base Representation (ABR) Number System, which retains the same range and storage density as binary for any given bit length, $n$.

The ABR number system introduces dynamic base adjustments, enabling distinct and optimized representations for each integer within the binary system’s scope. Through rigorously defined theorem and formal mathematical proofs, this work establishes the equivalence between the ABR and binary number systems, affirming that any number representable with $n$ bits in binary can also be represented with $n$ bits in ABR. Additionally, this system maintains a unique representation for each number, crucial for consistent data processing.

Beyond its theoretical underpinnings, ABR holds practical potential in digital steganography, where alternative encoding schemes could enhance data hiding and retrieval techniques. This paper explores the mathematical foundations, properties, and applications of ABR, opening new avenues in both theoretical number systems and practical computational designs.

\section{Background and Motivation}
The field of digital computation has been dominated by the binary number system due to its simplicity and reliability. This two-state system (0 and 1), is particularly suited to the on-off nature of electronic components and has become the default number system for data representation, storage, and transmission in digital computers. 

Creating an alternative number system that achieves at least the same storage density as the binary number system for $n$ bits while representing the same sets of numbers has proven to be a significant challenge. Many such systems either over-represent certain numbers while omitting others or represent a different range of numbers than the binary number system for the same number of bits. 

Other existing number systems such as mixed-radix ~\cite{wei2005number} and multi-base systems have different base values and real-world applications in clocks and calendars. Similarly, the double base number system ~\cite{doche2009double} has applications in multi-scalar multiplication. However, when implemented in digital computers for data representation, these systems do not achieve the same storage efficiency as the binary number system for the same number of bits.

This paper presents the Adaptive Base Representation (ABR) Number System, a novel alternative to the traditional binary number system used in digital computers. ABR dynamically adjusts base values based on specific criteria, offering new insights into number theory and digital encoding. By redefining base calculations and demonstrating their equivalence to binary ranges, this work aims to broaden understanding of numerical representation and its impact on computational design.
\section{Adaptive Base Representation Number System}
Assume a sequence of $n$ bits $S_a = (d_{n-1}, d_{n-2},...d_2, d_1, d_0)$ used to represent number in ABR Number System.
\subsection{Base Calculation}
Let \( B_i \) denote the base at index \( i \) in the Adaptive Base Representation System. The base values \( B_i \) are calculated as follows:
\[
    B_i = 
    \begin{cases} 
    i+2 & \text{if } i \in \{0,1\} \text{ and } n > 1. \\
    2^{i+1} - 1 - \sum_{\substack{j \text{ is odd}\\ \text{and } j < i}} B_j & \text{Otherwise}.
    \end{cases}
\]

\subsection{Representation Formula}

A number \( v \) represented in Adaptive Base Representation (i.e. ABR) Number System using \( n \) bits with sequence $S_a$ is expressed as:

\[
v = \sum_{i=0}^{n-1} (-1)^{\epsilon(i)} d_i \times B_i
\]

Where:

\[ d_i \in \{0, 1\} \text{ is the digit at index }i \text{ in sequence }S_a.\]

and, 

\[ \epsilon(i)= 
\begin{cases}    
1 & \text{if } i \text{ is even, and } i \leq n - 2, \text{and } d_{i} = 1, \text{ and } d_{i+1} = 1. \\
0 & \text{Otherwise}.
\end{cases}
\]
\section{Adaptive Base Representation Theorem and its Proof}
\begin{theorem}
For a given positive integer \( n \), the set of numbers representable with \( n \) bits in the Binary Number System is identical to the set of numbers representable with \( n \) bits in the Adaptive Base Representation Number System.
\\\\
Formally:

Let \( \mathbb{B}_n \) denote the set of numbers representable using \( n \) bits in the Binary Number System, and \( \mathbb{A}_n \) denote the set of numbers representable using \( n \) bits in the Adaptive Base Representation Number System, where $n$ is a positive integer. Then,

\[
\mathbb{B}_n = \mathbb{A}_n, \forall n \geq 1
\]
\end{theorem}
where:

\begin{itemize}
    \item \textbf{Binary Number System Set \( \mathbb{B}_n \):}
    
    The sequence $S_b = (b_{n-1}, b_{n-2},...b_2, b_1, b_0)$ represents binary number with $n$ bits.
    \[
    \mathbb{B}_n = \left\{ v \mid v = \sum_{i=0}^{n-1} b_i \times 2^i, \text{ where } b_i \in \{0, 1\} \text{ for } i = 0, 1, \ldots, n-1 \right\}
    \]

    \item \textbf{Adaptive Base Representation Number System Set \( \mathbb{A}_n \):}

    The sequence $S_a = (d_{n-1}, d_{n-2},...d_2, d_1, d_0)$ represents adaptive base representation number with $n$ bits.
    \[
    \mathbb{A}_n = \left\{ v \mid v = \sum_{i=0}^{n-1} (-1)^{\epsilon(i)} d_i \times B_i, \text{ for } i = 0, 1, \ldots, n-1 \right\}
    \]
    where,
    
    \[
    B_i = 
    \begin{cases} 
    i+2 & \text{if } i \in \{0,1\} \text{ and } n > 1. \\
    2^{i+1} - 1 - \sum_{\substack{j \text{ is odd}\\ \text{and } j < i}} B_j & \text{Otherwise}.
    \end{cases}
    \]

    and, 

    \[ d_i \in \{0, 1\} \text{ is the digit at index } i, \text{ The index is read from right to left in the sequence } S_a.\]
\\\\
    and, 
    
    \[ \epsilon(i)= 
    \begin{cases}
    1 & \text{if } i \text{ is even, and } i \leq n - 2, \text{and } d_{i} = 1, \text{ and } d_{i+1} = 1. \\
    0 & \text{Otherwise}.
    \end{cases}
    \]    
\end{itemize}

\begin{proof}
We need to show that the set of integers representable with \( n \) bits in the Binary Number System (BNS) is identical to the set of integers representable with \( n \) bits in the Adaptive Base Representation (ABR) Number System. Formally, \( \mathbb{B}_n = \mathbb{A}_n, \forall n \geq 1\).
\\
\\
\textbf{Case 1: (\( n = 1 \)):}
\\
For \( n = 1 \), in BNS:
\[
\mathbb{B}_1 = \{b_0 \times 2^0 \mid b_0 \in \{0, 1\}\} = \{0, 1\}
\]
For \( n = 1 \), in ABR:
\\
\[
\mathbb{A}_1 = \{d_0 \times B_0 \mid d_0 \in \{0, 1\}\}
\]
\\
The value obtained for $B_0$ when $n=1$ using $2^{i+1} - 1 - \sum_{\substack{j \text{ is odd}\\ \text{and } j < i}} B_j$ is,
\[
B_0 = 2^{0+1} - 1 = 1
\]
Thus when n=1, we have:
\[
\mathbb{A}_1 = \{d_0 \times 1 \mid d_0 \in \{0, 1\}\} = \{0, 1\}
\]

Hence, \( \mathbb{B}_1 = \mathbb{A}_1 \).
\\\\
\textbf{Case 2: (\( n = 2 \)):}
\\
For \( n = 2 \), in BNS:
\[
\mathbb{B}_2 = \{b_0 \times 2^0 + b_1 \times 2^1 \mid b_0 \in \{0, 1\}, b_1 \in \{0, 1\}\}  = \{0, 1, 2, 3\}
\]
\\\\
For \( n = 2 \), in ABR:
\\\\
We have the bits sequence for ABR represented as $S_a = \{d_1, d_0\}$ and the base represented as $B_i = i+2 $. That is,
\\\\
$B_0 = 0+2 = 2$ and, 
\\
$B_1 = 1+2 = 3$.
\\
The set $\mathbb{A}_2$ can be mathematically expressed as:
\[
\mathbb{A}_2 = \{(-1)^{\epsilon(0)} d_0 \times B_0 + (-1)^{\epsilon(1)} d_1 \times B_1 \mid d_0 \in \{0, 1\}, d_1 \in \{0, 1\} \}
\]
After updating values of $B_0$ and $B_1$, the expression of $\mathbb{A}_2$ can be written as:
\[
\mathbb{A}_2 = \{(-1)^{\epsilon(0)} d_0 \times 2 + (-1)^{\epsilon(1)} d_1 \times 3 \mid d_0 \in \{0, 1\}, d_1 \in \{0, 1\} \}
\]
\\
Now, let us change the values of $d_1$ and $d_0$, to check the numbers that can be represented with n = 2 in ABR Number System. 
\\\\
When $d_1 = 0$ and $d_0 = 0$, then represented number $v$ is:\\
${v= (-1)^0 \times 0 \times 3 + (-1)^0 \times 0 \times 2 = 0}$\\\\
When $d_1 = 0$ and $d_0 = 1$ :\\
${v = (-1)^0 \times 0 \times 3 + (-1)^0 \times 1 \times 2 = 2}$\\\\
When $d_1 = 1$ and $d_0 = 0$ :\\
${v = (-1)^0 \times 1 \times 3 + (-1)^0 \times 0 \times 2 = 3}$\\\\
When $d_1 = 1$ and $d_0 = 1$ :\\
${v = (-1)^0 \times 1 \times 3 + (-1)^1 \times 1 \times 2 = 3 - 2 = 1}$\\\\
Thus, 
\[
\mathbb{A}_2 = \{0, 1, 2, 3\}
\]

Hence, \( \mathbb{B}_2 = \mathbb{A}_2 \).
\\\\
\textbf{Analysis of Base $B_i$ in ABR:}
\\\\
Let's first understand how base values in BNS are developed. Assuming that $BNS\_Base_i$ represents base value at $i$ index in BNS. Then, it is defined as:
    \[BNS\_Base_{i} = 2^{i+1} -1 -\sum_{\substack{j < i}} BNS\_Base_{j}
    \]

     Here, $\sum_{\substack{j < i}} BNS\_Base_{j}$ is subtracted because when all the previous bits $b_j$ are set to 1 in BNS it represents the maximum number representable by previous bits.
    \\\\
    Thus in the ABR Number System,  when $n \geq 2$, the value of $B_i$ should be obtained after subtracting the maximum representable value from the previous bits.
    
    A number \( v \) when represented in ABR Number System using \( n \) bits with sequence $S_a= \{d_{n-1}, d_{n-2},..., d_1, d_0\}$ is expressed as:
    \[
    v = \sum_{i=0}^{n-1} (-1)^{\epsilon(i)} d_i \times B_i
    \]
    
    In the above expression, for the same values of $d_i$ and $B_i$, the value of $v$ is maximum when $\epsilon(i) = 0$, That is when $i$ is an odd number. 
    \\\\
    Thus, when $n \geq 2$, the value of base \( B_i \) should be calculated by subtracting the base values \( B_j \) of previous bits, where \( j \) is an odd number and \( j < i \). The formula for \( B_i \) is given by:
    \[B_i = 2^{i+1} - 1 - \sum_{\substack{j \text{ is odd}\\ \text{and } j < i}} B_j
    \]
    \\
    Now, a generalised form of $B_i$ can be created after combining the Case 1, and 2. 
    \\
    \( B_i \) is defined as:
    \[
    B_i = \begin{cases} 
    i+2 & \text{if } i \in \{0, 1\}, \text{ and } n > 1. \\
    2^{i+1} - 1 - \sum_{\substack{j \text{ is odd} \\ \text{and } j < i}} B_j & \text{Otherwise}.
    \end{cases}
    \]
    The recursive structure of \( B_i \) ensures that each \( v \) in BNS can always be represented in ABR by appropriately adjusting the base values.
\\\\
\textbf{Case 3: ($n \geq 2$):}
\\\\
\textbf{Inductive Hypothesis:}
\\
Assume \( \mathbb{B}_k = \mathbb{A}_k \) for some \( k \geq 2 \).
\\\\
\textbf{Inductive Step (\( n = k + 1 \)):}
\\
Consider any \( v \in \mathbb{B}_{k+1} \) represented as:
\[
v = \sum_{i=0}^{k} b_i \times 2^i
\]
We need to show that \( v \) can be represented in ABR with \( k+1 \) bits:
\[
v = \sum_{i=0}^{k} (-1)^{\epsilon(i)} d_i \times B_i
\]
\\
Since we have assumed that \( \mathbb{B}_k = \mathbb{A}_k \) for some \( k \geq 2 \), Now we need to showcase that for $n=k+1$ the base value $B_k$ holds \(\mathbb{B}_{k+1} = \mathbb{A}_{k+1}\).
We calculate \( B_k \) as follows:
    \[
    B_{k} = 
    2^{k+1} - 1 - \sum_{\substack{j \text{ is odd}\\ \text{and } j < k}} B_j 
    \]
This calculation supports that for \( n = k + 1 \), \( \mathbb{B}_{k+1} = \mathbb{A}_{k+1} \), completing the inductive step.
\\\\
\textbf{Proof of Representation Equivalence:}
\begin{itemize}
    \item \textbf{Base Values \( B_i \):} 
    \\
    For \( n = k + 1 \), The base values \( B_i \) are defined as:

    \[
    B_i = 
    \begin{cases} 
    i+2 & \text{if } i \in \{0,1\} \text{ and } k+1 > 1. \\
    2^{i+1} - 1 - \sum_{\substack{j \text{ is odd}\\ \text{and } j < i}} B_j & \text{Otherwise}.
    \end{cases}
    \]

    This definition is designed so that the sequence of bits $S_a$ with \( B_i \) base values can represent the highest number that the binary number system can represent with $k+1$ bits.

    \item \textbf{Corrective Term \( (-1)^{\epsilon(i)} \):} 
    \\
    The function \( \epsilon(i) \) adjusts the sign of certain terms in ABR:
    \[
    \epsilon(i) = \begin{cases}
    1 & \text{if } i \text{ is even, and } i \leq k - 1, \text{and } d_{i} = 1, \text{ and } d_{i+1} = 1. \\
    0 & \text{otherwise}.
    \end{cases}
    \]
    This adjustment ensures that consecutive 1's do not cause over-representation, maintaining a unique representation of bits when representing $v$ integer in the ABR number system.
    \\
    \item \textbf{Inductive Conclusion:} 
    \\
    Given the structure of \( B_i \) and the corrective term \( \epsilon(i) \), every integer \( v \) representable in \( \mathbb{B}_{k+1} \) can also be represented in \( \mathbb{A}_{k+1} \). Hence, \( \mathbb{B}_{k+1} = \mathbb{A}_{k+1} \).
\end{itemize}
\textbf{Conclusion:} 
\\
We conclude that \( \mathbb{B}_n = \mathbb{A}_n \) for all \( n \geq 1 \). Thus, the sets of numbers representable in the Binary Number System and the Adaptive Base Representation Number System are indeed the same for $n$ number of bits. This equivalence also ensures that each number \( v \) represented in the ABR number system will have a unique bit representation. 

Hence, the proposed ABR number system can be an alternative to the Binary number system in digital computers for data representation, storage, and transmission operations.
\end{proof}
\section{In-depth analysis of ABR Number System}
In this section, we analysed the base values of ABR and compared them with those of BNS. Additionally, we compared the 4-bit encodings of numbers in the ABR and BNS number systems.

\subsection{Base value comparison in BNS and ABR}
Let's first analyse the base values for both ABR and BNS number systems when $n=16$. We have shown the comparison in Table~\ref{tab1}.

\begin{table}
\centering
\caption{Comparison of base values in ABR and Binary number systems,
$n$ = 16.}\label{tab1}
\begin{tabular}{|c|c|c|}
\hline
Bit Index ($i$) &  Base in BNS & Base in ABR\\
\hline
0 & 1 & 2  \\
1 & 2 & 3  \\
2 & 4 & 4  \\
3 & 8 & 12  \\
4 & 16 & 16  \\
5 & 32 & 48  \\
6 & 64 & 64  \\
7 & 128 & 192  \\
8 & 256 & 256  \\
9 & 512 & 768  \\
10 & 1024 & 1024 \\
11 & 2048 & 3072 \\
12 & 4096 & 4096 \\
13 & 8192 & 12288 \\
14 & 16384 & 16384 \\
15 & 32768 & 49152 \\
\hline
\end{tabular}
\end{table}
From Table~\ref{tab1} it can be easily established that when $n > 1 \text{ and } i > 1$ then the base value in ABR and BNS number systems are equal for all the even index values (i.e. when $i$ is an even number).
\subsection{Number Representation Examples in ABR}
We have also illustrated a comparison between the binary number system, and the ABR number system for representing integers from 0 to 15 using 4 bits in Table~\ref{tab2}. In this example, $n = 4$, and the base values in ABR Number System are $B_3 = 12$, $B_2 = 4$, $B_1 = 3$, and $B_0 = 2$.
\begin{table}
\centering
\caption{Comparison of Decimal, Binary, and Adaptive Base Representation Number Systems for values from 0 to 15, using 4 bits.}\label{tab2}
\centering
\begin{tabular}{|c|c|c|}
\hline
Decimal Number & Binary Number & ABR Number \\
\hline
0 & 0000 & 0000  \\
1 & 0001 & 0011  \\
2 & 0010 & 0001  \\
3 & 0011 & 0010  \\
4 & 0100 & 0100  \\
5 & 0101 & 0111  \\
6 & 0110 & 0101  \\
7 & 0111 & 0110  \\
8 & 1000 & 1100  \\
9 & 1001 & 1111  \\
10 & 1010 & 1101 \\
11 & 1011 & 1110 \\
12 & 1100 & 1000 \\
13 & 1101 & 1011 \\
14 & 1110 & 1001 \\
15 & 1111 & 1010 \\
\hline
\end{tabular}
\end{table}

\begin{figure}
\includegraphics[width=\textwidth]{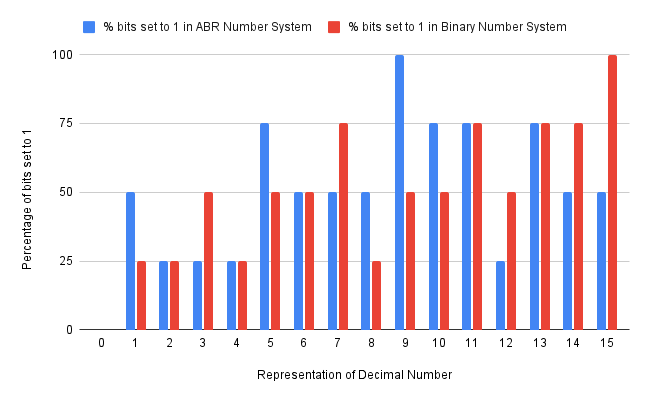}
\caption{Visual analysis of bit usage when representing numbers in ABR and Binary number systems.} \label{fig1}
\end{figure}

We have also provided a chart in Figure~\ref{fig1} for visually understanding the impact on bit usage (i.e. bit set to 1) when representing a large decimal number in ABR and BNS number systems.

One can see in Figure \ref{fig1} that the ABR number system has maximum bit usage (i.e. when all bits are set to 1) to approximately represent a middle region number, while in the BNS the maximum bit usage is when it is representing the highest number (i.e. $2^n-1$).
\section{Applications of Adaptive Base Representation Number System}
\subsection{In Digital Computers}
\begin{itemize}
    \item \textbf{Data Storage and Transmission:} ABR number system provides a new scheme of uniquely representing data thus providing a new consistent encoding scheme. Thus, it can be used for data storage ~\cite{mariani2021explaining}, and transmission activities. 
    \item \textbf{Steganography in Cyber Security:} Combining BNS with ABR Number System can provide a powerful steganography mechanism ~\cite{csahin2021review,cheddad2010digital} for representing data in bits. An example of implementing steganography is shown in Figure~\ref{fig2}.
    \begin{figure}
    \includegraphics[width=\textwidth]{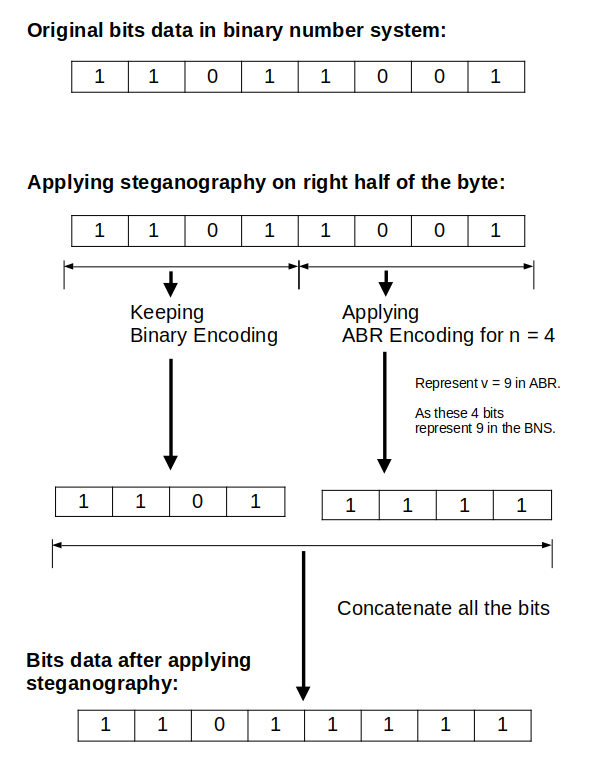}
    \caption{Using ABR with Binary number system for implementing steganography in 1 byte data.} \label{fig2}
    \end{figure}
    \\
    The conventional 8-bit binary data representation can be altered to convert 4 bits into their corresponding Adaptive Base Representation (ABR) encoding. Consequently, without identifying which bits correspond to the binary encoding and which represent the ABR encoding, it becomes difficult to recover the original data.
    \item \textbf{Error Detection and Correction:} The ABR number system is fully compatible with existing error detection and correction methods, such as Hamming codes \cite{6772729}. An example illustrating the application of Hamming codes to detect and correct a single-bit error for the decimal number 13 represented in BNS and ABR number system is provided. The process of calculating parity bits for the number 13 using the even parity is shown in Table \ref{tab3}. Additionally, Table \ref{tab4} demonstrates the error correction mechanism, assuming the value of leftmost bit (i.e. $d_4$) is altered during transmission.

\begin{table}
\centering
\caption{Calculation of parity bits for Hamming code with even parity in both BNS and ABR number systems for the decimal number 13.}\label{tab3}
\centering
\begin{tabular}{|c|c|c|c|}
\hline
Step & Description & Binary Number System & ABR Number System\\
\hline
1 & Represent 13 in data bits & 1 1 0 1 & 1 0 1 1  \\
2 & Order of data bits & $d_4 \text{ } d_3 \text{ } d_2 \text{ } d_1$ & $d_4 \text{ } d_3 \text{ } d_2 \text{ } d_1$ \\
3 & Position of parity and data bits & $d_4 \text{ } d_3 \text{ } d_2 \text{ } p_3 \text{ } d_1 \text{ } p_2 \text{ } p_1$ & $d_4 \text{ } d_3 \text{ } d_2 \text{ } p_3 \text{ } d_1 \text{ } p_2 \text{ } p_1$\\
4 & Bit level index & 7, 6, 5, 4, 3, 2, 1 & 7, 6, 5, 4, 3, 2, 1 \\
5 & Parity and data bits & $1 \text{ } 1 \text{ } 0 \text{ } p_3 \text{ } 1 \text{ } p_2 \text{ } p_1$ & $1 \text{ } 0 \text{ } 1 \text{ } p_3 \text{ } 1 \text{ } p_2 \text{ } p_1$\\
6 & Value of $p_1$ & 0 & 1  \\
7 & Value of $p_2$ & 1 & 0  \\
8 & Value of $p_3$ & 0 & 0  \\
9 & Parity and data bits & 1 1 0 0 1 1 0 & 1 0 1 0 1 0 1  \\
\hline
\end{tabular}
\end{table}

\begin{table}
\centering
\caption{Demonstrating error correction with even parity when the leftmost bit (i.e. $d_4$) of the decimal number 13 is altered during transmission.}\label{tab4}
\centering
\begin{tabular}{|c|c|c|c|}
\hline
Step & Description & Binary Number System & ABR Number System \\
\hline
1 & Received parity and data bits & 0 1 0 0 1 1 0 & 0 0 1 0 1 0 1 \\
2 & Position of parity and data bits & $d_4 \text{ } d_3 \text{ } d_2 \text{ } p_3 \text{ } d_1 \text{ } p_2 \text{ } p_1$ & $d_4 \text{ } d_3 \text{ } d_2 \text{ } p_3 \text{ } d_1 \text{ } p_2 \text{ } p_1$\\
3 & Bit level index & 7, 6, 5, 4, 3, 2, 1 & 7, 6, 5, 4, 3, 2, 1 \\
4 & Syndrome bit $s_1$ depends on & $d_4 \text{, } d_2 \text{, } d_1 \text{, } p_1$ & $d_4 \text{, } d_2 \text{, } d_1 \text{, } p_1$ \\
5 & Calculate syndrome bit $s_1$ & 1 & 1 \\
6 & Syndrome bit $s_2$ depends on & $d_4 \text{, } d_3 \text{, } d_1 \text{, } p_2$ & $d_4 \text{, } d_3 \text{, } d_1 \text{, } p_2$ \\
7 & Calculate syndrome bit $s_2$ & 1 & 1 \\
8 & Syndrome bit $s_3$ depends on & $d_4 \text{, } d_3 \text{, } d_2 \text{, } p_3$ & $d_4 \text{, } d_3 \text{, } d_2 \text{, } p_3$ \\
9 & Calculate syndrome bit $s_3$ & 1 & 1 \\
10 & Error in bit position  & 7 & 7 \\
11 & Updated parity and data bits & 1 1 0 0 1 1 0 & 1 0 1 0 1 0 1 \\
12 & Corrected data bits & 1 1 0 1 & 1 0 1 1 \\
13 & Do the data bits represent 13? & YES & YES \\
\hline
\end{tabular}
\end{table}
    \item \textbf{Data Compression:} For base-dependent compression algorithms, the Adaptive Base Representation (ABR) number system can represent any decimal number that the binary number system can, using the same number of bits, by converting binary to decimal and storing its ABR equivalent representation. In base-independent data compression algorithms, like Huffman \cite{4051119} or arithmetic coding, bit sequences can be directly used in the ABR number system, avoiding conversions and reducing storage overhead. Further optimizations in existing data compression algorithms can also be explored based on the unique properties of the ABR number system.
    \item \textbf{Arithmetic Operations:} Potential optimization in arithmetic circuits can be achieved by reducing carry propagation delays ~\cite{funk2024binary}.
\end{itemize}

\subsection{In Other Fields}
\begin{itemize}
    \item \textbf{Information Theory:} ABR's variable base system can aid in designing encoding schemes ~\cite{witten2020mini} that maximize information density ~\cite{fabris2009shannon} for non-uniform data distributions.
    \item \textbf{Embedded Systems:} Optimizing bit usage for resource-constrained environments where memory and processing power are limited ~\cite{marwedel2021embedded}.
\end{itemize}

\subsection{Suggestive Application in Mathematics}
\textbf{Number Theory:} The ABR number system offers potential applications in the study of integer sequences and representation problems ~\cite{landau2021elementary}. It may provide new insights into the distribution of primes and other number-theoretic functions by enabling alternative representations and analyses of integer sequences. This novel approach could also contribute to advancements in factorization and modular arithmetic ~\cite{muller2005analysis}.
\section{Conclusion}
The proposed Adaptive Base Representation (ABR) Theorem demonstrates that both ABR and Binary Number Systems (BNS) represent the same sets of numbers using $n$ number of bits. It guarantees that each decimal number $v$ has a unique bit representation in the ABR number system. Consequently, the ABR theorem and its associated number system offer a viable alternative to the binary number system for data representation, storage, and transmission in digital computers. The ABR number system is also compatible with existing data compression algorithms like Huffman coding and arithmetic coding, as well as error detection and correction mechanisms such as Hamming codes. Integrating ABR with the binary number system can create a robust steganography mechanism for digital data. Additionally, the ABR number system presents an exciting framework for exploring new methodologies in number theory and bit encoding in digital computers.
\begin{credits}
\subsubsection*{Funding.}
No Institutional funding is received to perform this research work.
\subsubsection{\discintname}
The authors have no competing interests to declare that are relevant to the content of this article.
\end{credits}
\bibliographystyle{splncs04}
\bibliography{main}
\end{document}